\documentclass[english,twocolumn]{article}

\usepackage[utf8]{inputenc}
\usepackage[big]{dgruyter}
\usepackage{nicefrac}
\usepackage{gensymb}

\setcitestyle{authoryear,round}
  
\begin{document}
 
  \articletype{Research Article{\hfill}Open Access}

  \author*[1]{Avdeeva Aleksandra}

\author[2]{Kovaleva Dana}
\author[3]{Malkov Oleg}
\author[4]{Nekrasov Alexey}


  \title{\huge Fitting procedure for estimating interstellar extinction at high galactic latitudes}

  \runningtitle{Procedure for interstellar extinction estimation}


  \begin{abstract}
{We determine the interstellar extinction in the selected
high-latitude areas of the sky based on Gaia EDR3 astrometry and
photometry and spectroscopic data from RAVE survey. We approximate the
results with the cosecant law in each area thus deriving the
parameters of the barometric formula for different lines of sight. The
distribution of the parameters over the entire sky is described using
spherical harmonics. As a result, we get a mathematical description of
the interstellar visual extinction for different lines of sight and
distances from the Sun which can be used for estimating interstellar
extinction.}
\end{abstract}

  \journalname{Open Astronomy}
\DOI{DOI}
  \startpage{1}
  \received{..}
  \revised{..}
  \accepted{..}

  \journalyear{..}
  \journalvolume{..}

\maketitle


{ \let\thempfn\relax
\footnotetext{\hspace{-1ex}{\Authfont\small \textbf{Corresponding Author: Avdeeva Aleksandra:}} {\Affilfont Institute of Astronomy of the Russian Acad. Sci., Moscow 119017, Russia - HSE University, Moscow 101000, Russia, Email: avdeeva@inasan.ru}}
}

{ \let\thempfn\relax
\footnotetext{\hspace{-1ex}{\Authfont\small \textbf{Kovaleva Dana:}} {\Affilfont Institute of Astronomy of the Russian Acad. Sci., Moscow 119017, Russia}}
}

{ \let\thempfn\relax
\footnotetext{\hspace{-1ex}{\Authfont\small \textbf{Malkov Oleg:}} {\Affilfont Institute of Astronomy of the Russian Acad. Sci., Moscow 119017, Russia}}
}

{ \let\thempfn\relax
\footnotetext{\hspace{-1ex}{\Authfont\small \textbf{Nekrasov Alexey:}} {\Affilfont Physics Faculty of Lomonosov Moscow State University, Moscow 119992, Russia - Sternberg Astronomical Institute Lomonosov Moscow State University, Moscow
119992, Russia}}
}

\section{Introduction}

Taking extinction into account is an essential step in almost every astronomical, and particularly astrophysical, research. For this reason, 3D interstellar extinction maps are tools of great usage. Moreover, the distribution of interstellar dust, which is mostly responsible for an extinction, is itself interesting in a context of Galaxy evolution and structure studies. \\
Although there are numerous approaches to a 3D mapping, the results are usually presented in a form of a discrete set of extinction values corresponding to different distances from the Sun in a certain direction on the celestial sphere. \cite{Marshall2006} and \cite{Robin2015} imply the population synthesis Besançon model of the Galaxy along with 2MASS and WISE photometry. The 3D map constructed is available as a table in Vizier. \cite{Sale2014} produced a 3D map based on hierarchical Bayesian model using the IPHAS photometric data. \cite{Green2015} and \cite{Green2018} also constructed a 3D map using the Bayesian method, SDSS photometry at first and Pan-STARRS 1 and 2MASS photometry later. The results can be downloaded for the particular lines of sight or en bloc for all available directions. \cite{Lallement2014} implied regularized Bayesian approach to color excesses datasets based on Strömgren catalogs, the Geneva photometric database, and the Geneva-Copenhagen survey. Later, \cite{Lallement2019} used 2MASS photometry and Gaia DR2 astrometry to produce a map of interstellar dust within 3 kpc. Lately, \cite{Gontcharov2020} have built the model of interstellar extinction based on the Gaia DR2 parallaxes and Gaia and WISE photometry for over a hundred thousand of giants and the result is accessible on a request. \\
To sum up, despite the large number of approaches for building detailed maps, the result is often presented in the form of tables that one need to download and refer to every time needed. It would be much more convenient to use a formula, which can estimate the extinction for known direction and distance. \\ 
Possible solution was introduced in \cite{2021EPJST.tmp..185N}. Authors proposed an analytical description of the dependence of visual extinction $A_V$ on galactic latitude $b$ and distance $d$. Authors used the cosecant law:
\begin{equation}
    A_V(b,d) = \frac{a_0 \cdot \beta}{\sin|b|} \Big(1 - \exp \Big({\frac{-d \cdot \sin|b|}{\beta}\Big)}\Big), 
    \label{eq:parenago}
\end{equation}
 that was at first introduced by \cite{Parenago1940}. The parameter $\beta$ is the scale height, and $a_0$ is the extinction per unit length in the Galactic plane.
The work is based on the Gaia EDR3 photometry and spectroscopic parameters from LAMOST data. The details could be found in the original paper.\\
Following the \cite{2021EPJST.tmp..185N} we use Parenago formula. We estimated them individually in 40 different areas of the sky via the approximation of visual extinction $A_V$
calculated on the basis of RAVE DR6 spectroscopic data (\cite{McMillan2020}) and Gaia EDR3 photometry and astrometry (\cite{Prusti2016}, \cite{Brown2021}). Then we approximated the estimated in this work parameters along with the ones found by \cite{2021EPJST.tmp..185N} by spherical harmonics over the entire sky. \\
 The paper is organised as follows. In section 2 we describe the data that was used and calculate the visual extinction of objects at individual areas of the sky. Section 3 contains the description of approximation methods and the resulting parameters $a_0$, $\beta$. In section 4 we present the results of approximation over the entire sky. Finally, in section 5 we discuss the perspectives of this work.
 
 \section{Observational data and obtaining the extinction}\label{sec:data}

\begin{figure}
    \centering
    \includegraphics[width=\columnwidth]{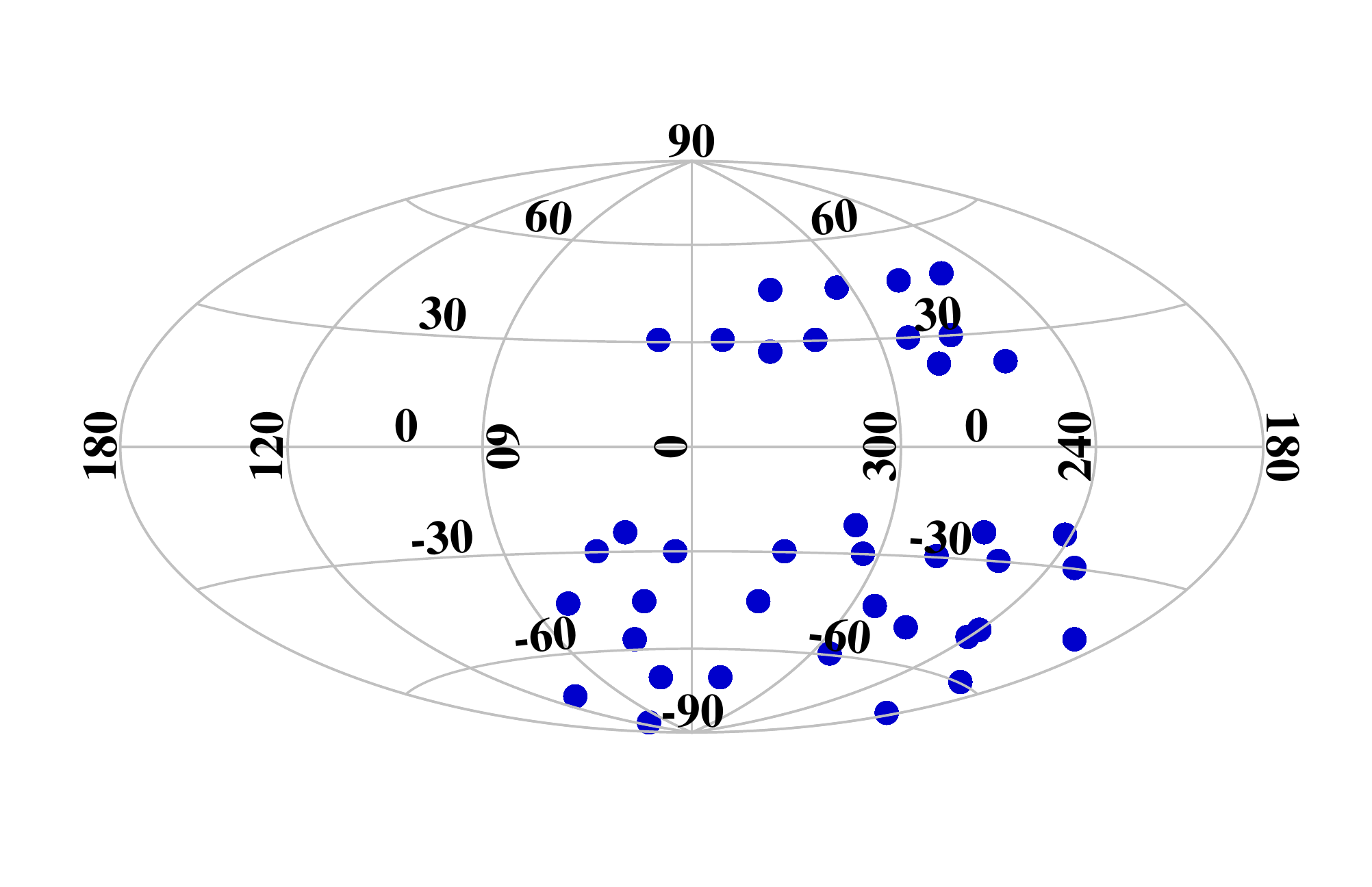}
    \caption{The distribution of selected areas over the sky in galactic coordinates. Aitoff projection.}
    \label{fig:fig1}
\end{figure}

\begin{figure}
    \centering
    \includegraphics[width=\columnwidth]{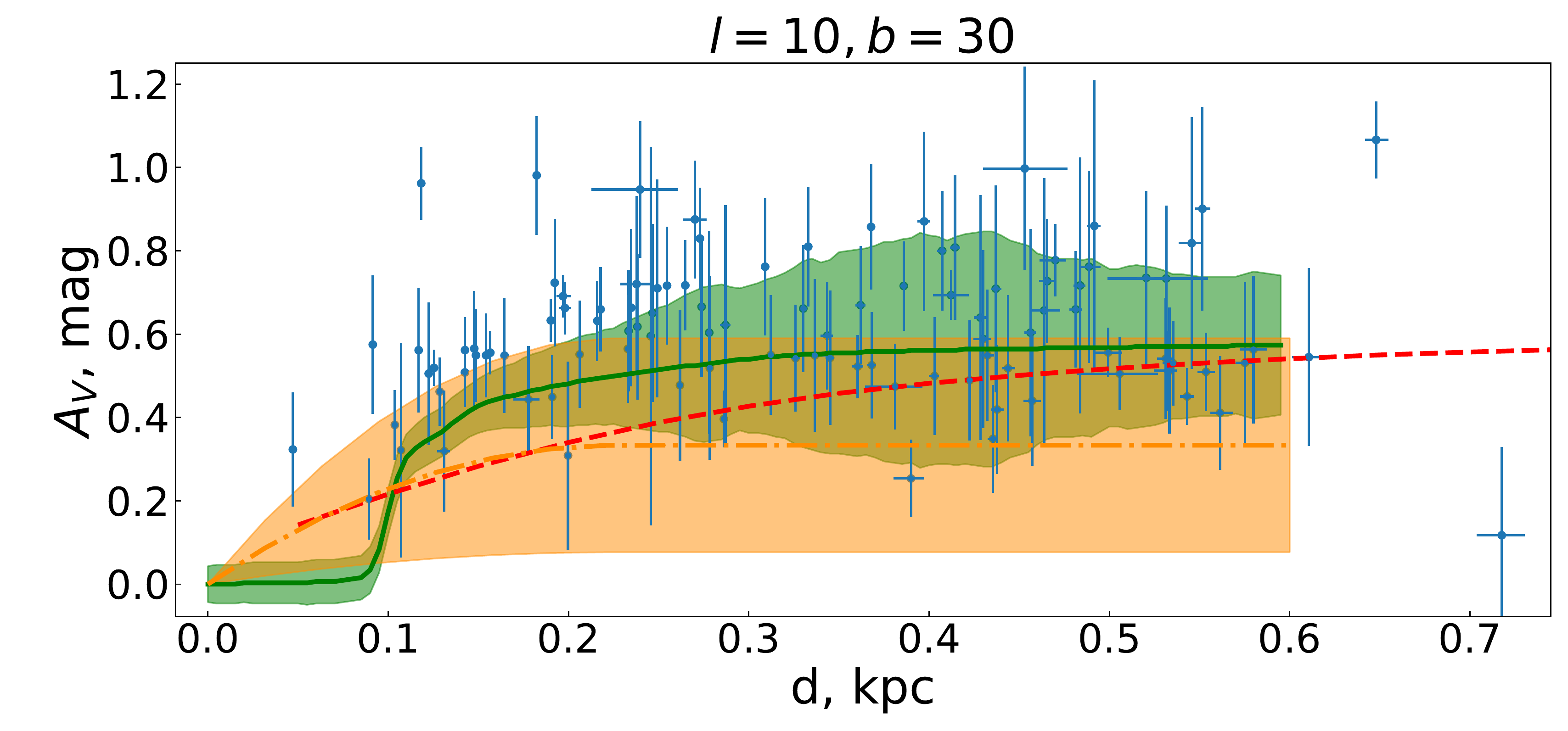} \\
    \includegraphics[width=\columnwidth]{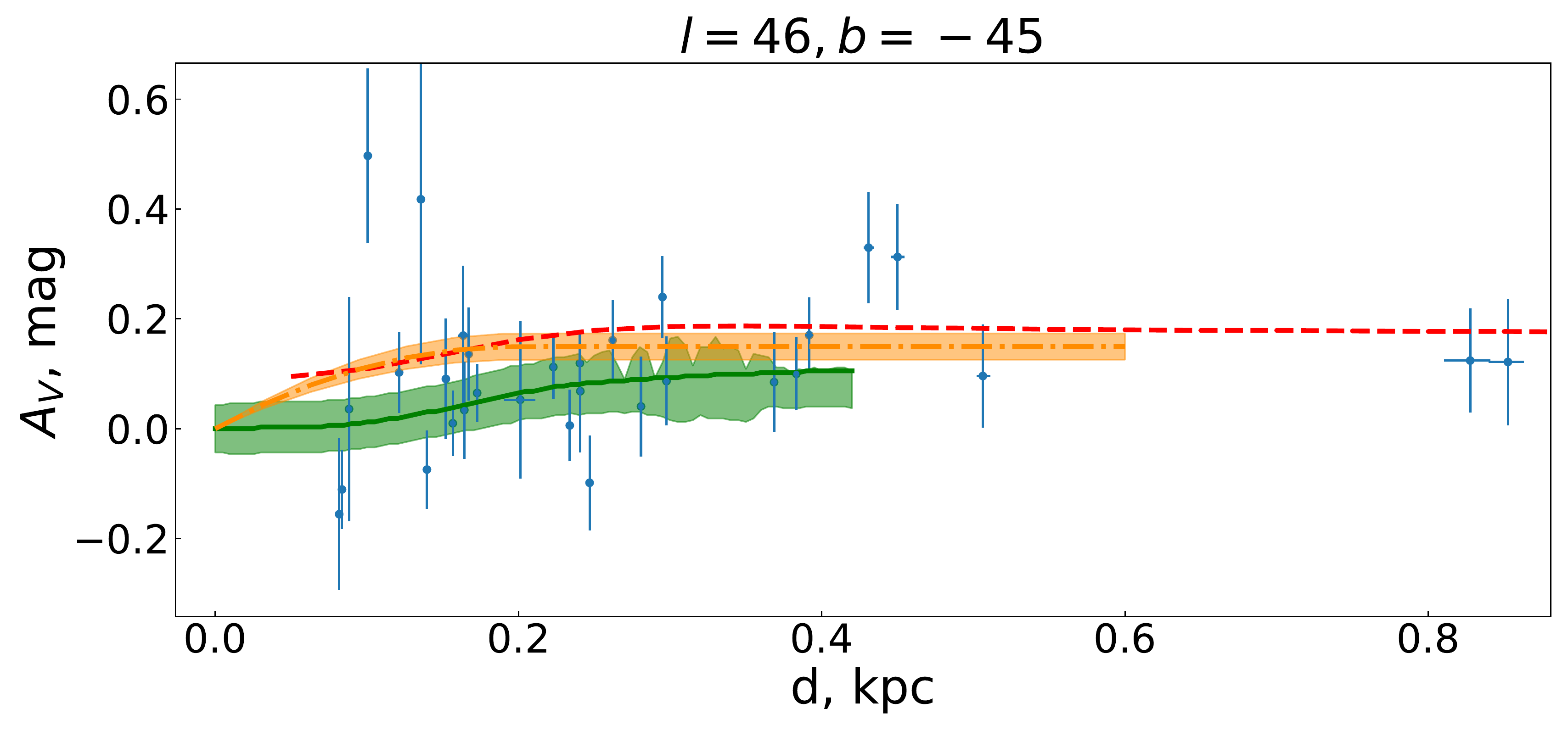} \\
    \includegraphics[width=\columnwidth]{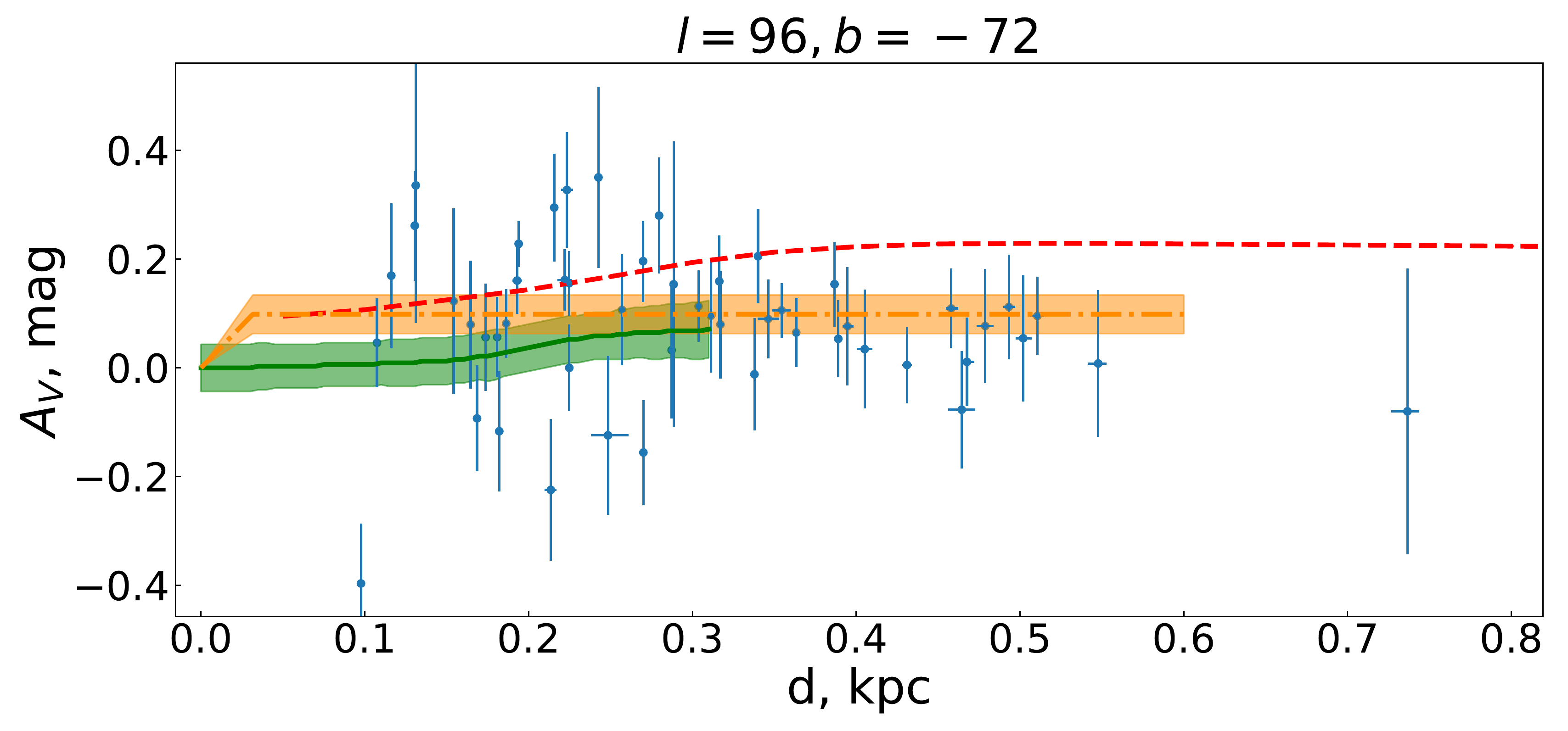} \\
    \includegraphics[width=\columnwidth]{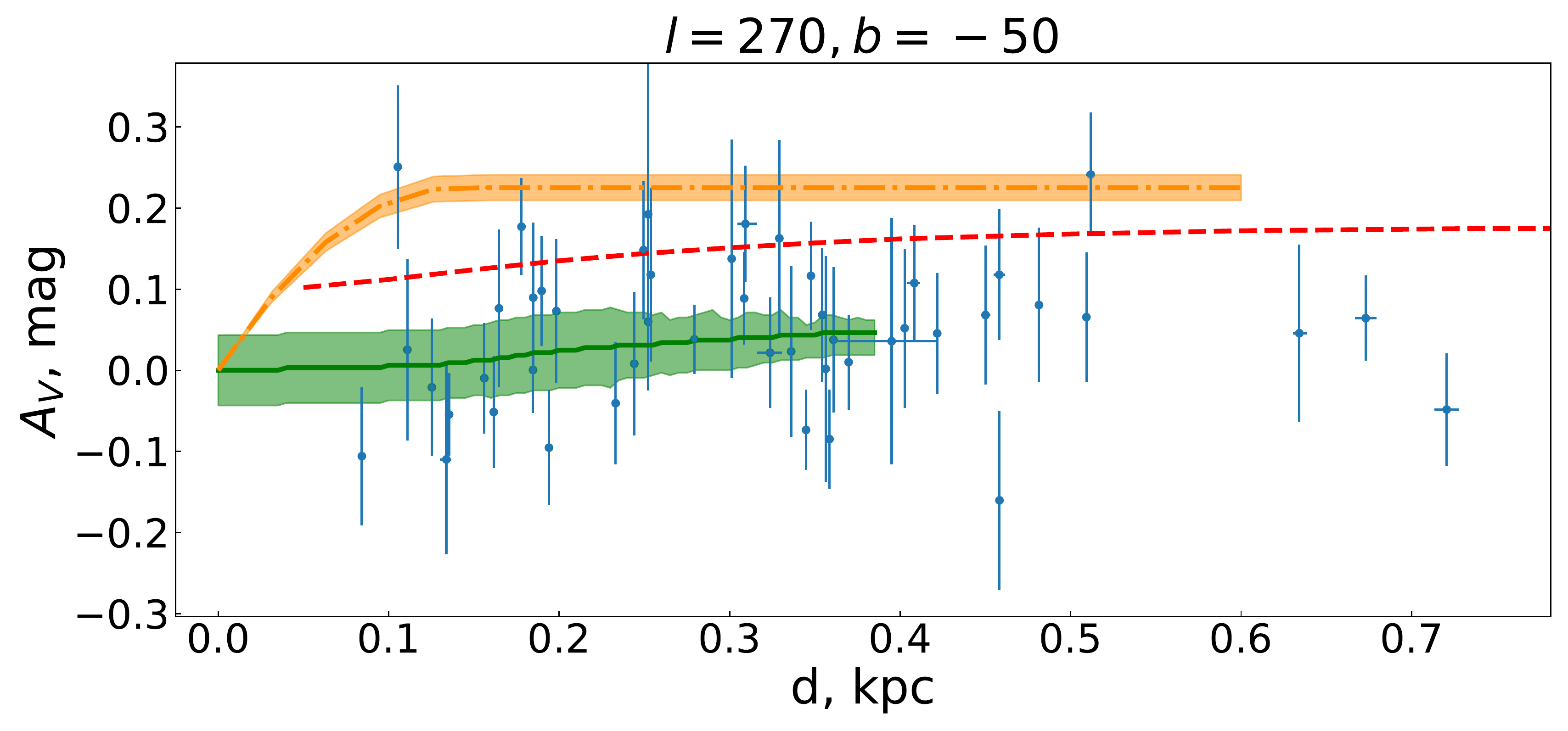} 
    \caption{Comparison of our extinctions with three different 3D maps or models of interstellar extinction. Green solid line is made from values provided by Stilism (\cite{Lallement2014}). Orange dash-dotted line adopted from \cite{Arenou1992}. Red dashed line refers to a \cite{Gontcharov2020} model, according to authors, uncertainties of $A_V$ are less than $0.03$ mag.} 
    \label{fig:fig2}
\end{figure}

We have selected 40 different areas in the RAVE coverage at high ($|b|>20^\degree$) latitudes (see Fig.\ref{fig:fig1}). A limit of $|b|$ was chosen because the cosecant law is believed to satisfactory reproduce the $A_V(d)$ dependence at high latitudes far from the Galactic plane. Each area is a cone of 80 arcsec radius, defined by its center.\\
We made a cross-match of RAVE DR6 and Gaia EDR3 objects within every area. For each object distances were adopted from \cite{BailerJones2021} catalogue. We have chosen BDASP catalogue of RAVE DR6 to get the spectroscopic parameters of stars. \\
For individual stars in each area we have calculated the visual extinction, using the following formula:
\begin{equation}
\label{eq:eq2}
    A_V= \frac{c1}{c2}\cdot\Big((BP-RP)-(BP-RP)_0\Big),
\end{equation}
where $c1$ is $A_G$ to $E(BP-RP)$ ratio and $c2$ is an absorption coefficient, $A_G$ to $A_V$ ratio. $A_G$ is the interstellar extinction in G-band of Gaia. According to Bono et.al c2 = 0.84. The coefficient c1 = 2.02 was calculated according to \cite{Cardelli1989} assuming that the $R_V$ value equals to 3.1. \\
In fact, $c2$ is not a constant due to broad G-band of Gaia.   \\
The intrinsic colours  $(BP-RP)_0$ were calculated according to \cite{Pecaut2013} (the table is published also on a \href{http://www.pas.rochester.edu/~emamajek/EEM_dwarf_UBVIJHK_colors_Teff.txt}{$\textit{web page}$}), where the intrinsic colours provided for different effective temperatures. For this purpose we have used the effective temperatures $T_{\rm eff}$ of RAVE DR6 BDASP catalogue. In the intervals between the values given in the table we performed a linear approximation. \\
In some areas the overall extinction trend with distance turned out to be negative due to low values of extinction to distant objects. We associate this with improper temperature derivation for some objects in RAVE data. To eliminate the stars with outrageously low extinctions at far distances, we set the following restrictions on the data used for the work: $1.6>BP-RP>0.8$ and $\log g>3.5$. Other data issues and possible reasons are discussed in the Sec. \ref{subsec:chi}. \\
We compared our extinctions remaining after sifting with three different models or 3D maps of interstellar extinction in four different areas of the sky. The comparison is presented in Fig \ref{fig:fig2}. 

\section{Deriving the parameters of the cosecant law in individual areas}  

To obtain the parameters $a_0$ and $\beta$ of the cosecant law we have minimized the following $\chi^2$ functional:

\begin{equation}
\label{eq:functional}
    \chi^2=\sum_{n=1}^N \Big(\frac{A_V(d_n)-A_{V,n}}{\epsilon_n(d_n,A_{V,n})}\Big)^2
\end{equation}
Here $A_{V,n}$ are the extinction values we have calculated in Sec. \ref{sec:data}; $A_V(d_n)$ - values, predicted by the cosecant law (\ref{eq:parenago}) for a distance $d_n$; $\epsilon_n$ - uncertainty of extinction and distances, N is the total number of objects in the area. \\
We used two methods for finding the minimum of the functional: beat-fit $\chi^2$ minimization and $\chi^2$ scan. Methods are described below.

\subsection{Best-fit $\chi^2$ minimization} \label{subsec:chi}
\begin{figure}
	\begin{minipage}[h]{0.49\linewidth}
		\center{\includegraphics[width=\textwidth]{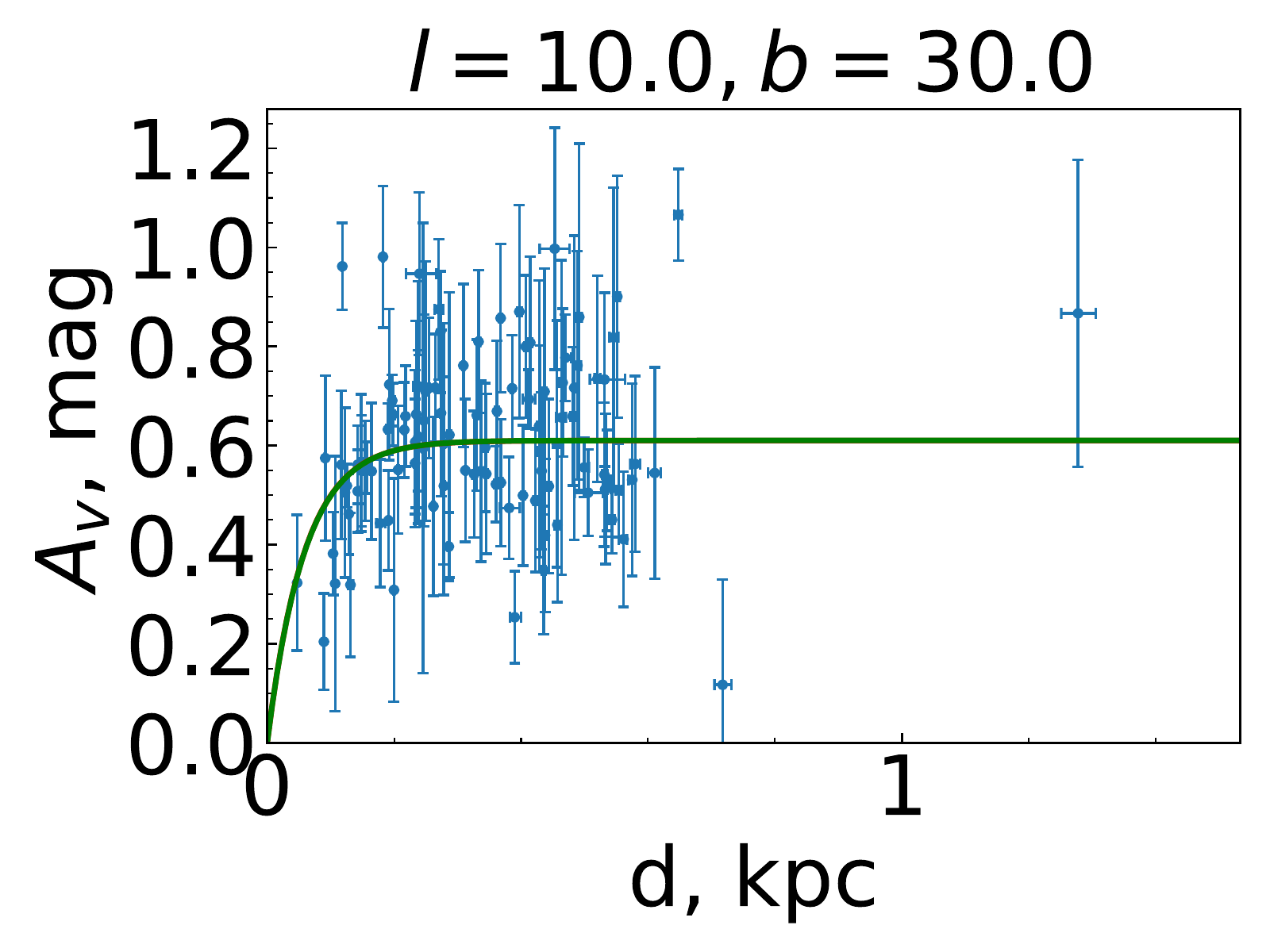}}
	\end{minipage}
	\hfill
	\begin{minipage}[h]{0.49\linewidth}
		\center{\includegraphics[width=\textwidth]{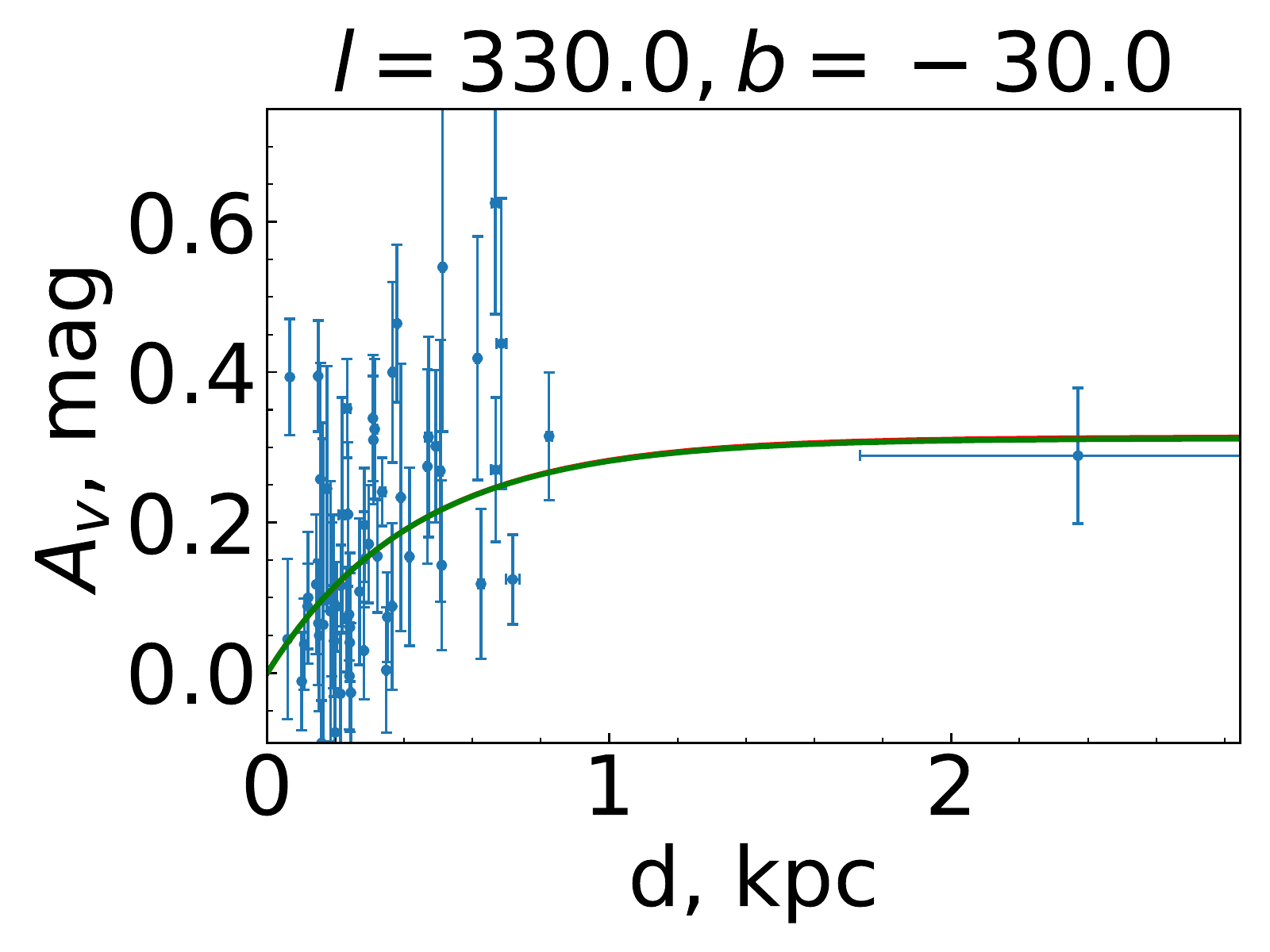}}
	\end{minipage}
	\vfill
	\begin{minipage}[h]{0.49\linewidth}
		\center{\includegraphics[width=\textwidth]{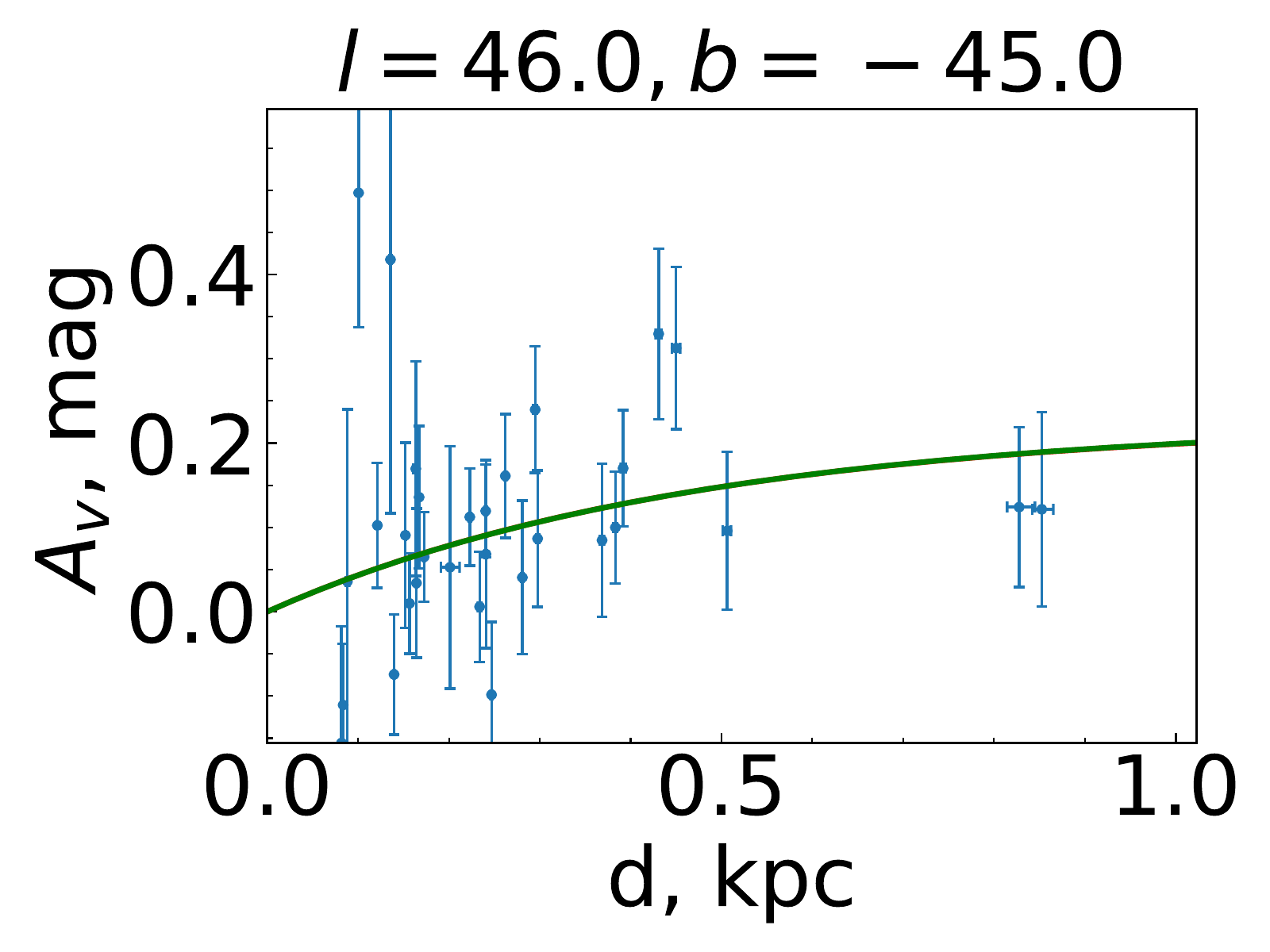}}
	\end{minipage}
	\hfill
	\begin{minipage}[h]{0.49\linewidth}
		\center{\includegraphics[width=\textwidth]{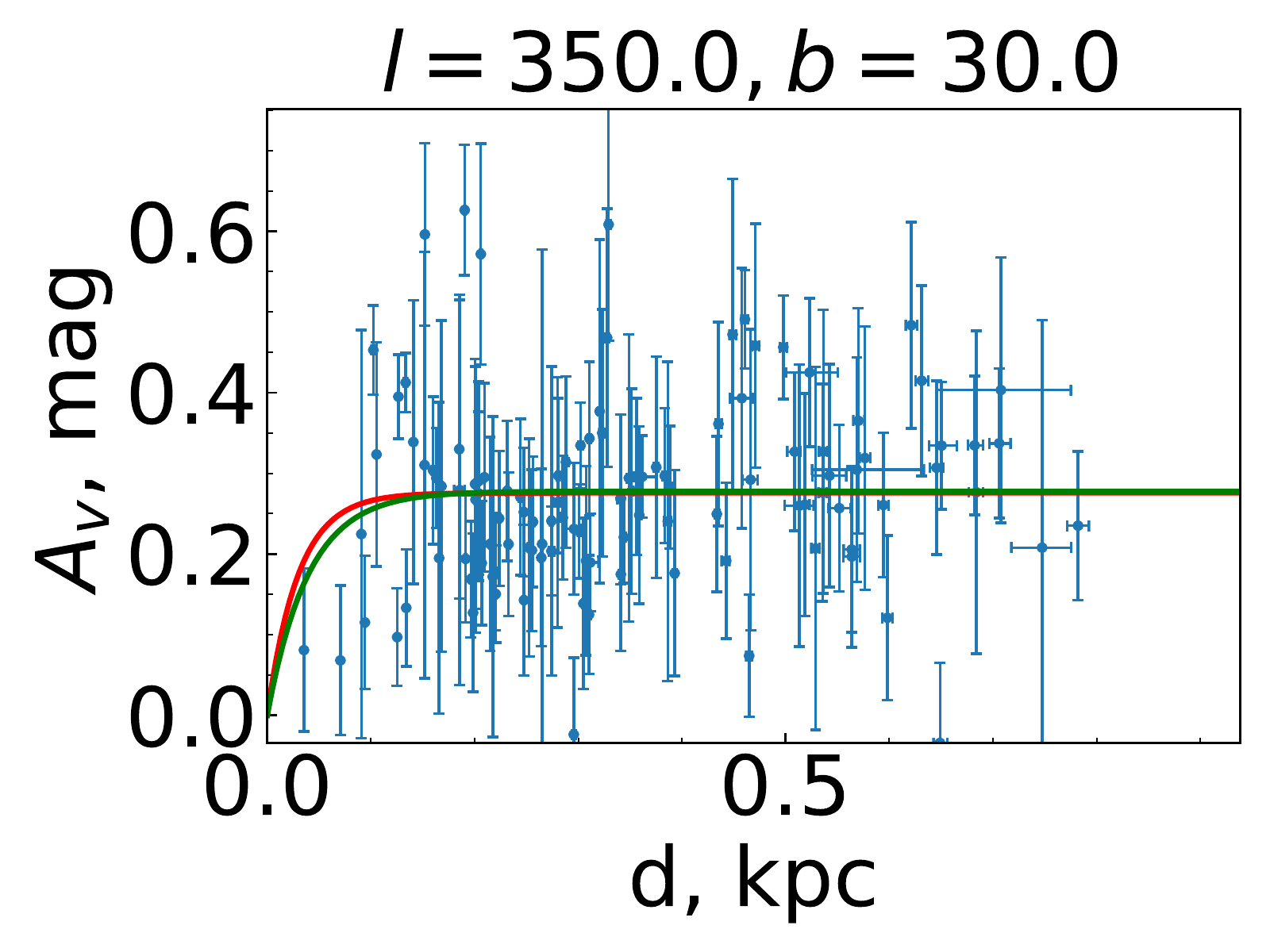}}
	\end{minipage}
	\caption{Examples of $\chi^2$ minimization. Galactic coordinates of the center of the region are on top of each graph. Green line is the result of approximation with best-fit $\chi^2$ minimization and the red one is the solution obtained with $\chi^2$-scan.}
    \label{fig:apprx1}
\end{figure} 
The minimization of Eq.(\ref{eq:functional}) was performed with lmfit package of Python \citep{Newville2014}. We have used the Levenberg-Marquardt algorithm. The standart errors are $1\sigma$ uncertainty, estimated by inverting the Hessian matrix, calculated also with lmfit package. Examples of successful estimation of $a_0$ and $\beta$ parameters are presented in Fig. \ref{fig:apprx1}.      \\
The approximation was performed successfully for 36 of 40 areas. In three other areas there was no solution found. It could be both due to the number of negative values of extinction and the inadequate trend of extinction, non-monotonic or decreasing. Another one area we have eliminated is the one with doubtfully low value of $a_0$. We have kept the solutions with preposterous values of uncertainties such as areas number 3, 20 and 37 because they have reasonable values of $a_0$ and $\beta$ and thus can contribute to the all-sky approximation.\\
We assume the main problem with approximation in "bad" regions is due to improper values of $T_{\rm eff}$ in RAVE catalogue. We believe so since the temperatures adopted from the LAMOST data and the RAVE data for the same objects from the intersection of the surveys differ systematically. While the research of \cite{2021EPJST.tmp..185N} based on the LAMOST data did not show this strong discrepances with the cosecant law, our present research based on the RAVE DR6 data, sometimes shows the result inconsistent with Eq.(\ref{eq:parenago}). The offsets in LAMOST–RAVE overlapping data have been noticed earlier \cite{Vickers2018} for the radial velocities. The comprehensive analysis of temperatures offset will be made and published in a subsequent paper.     

\subsection{$\chi^2$ scan}

\begin{figure}
	\begin{minipage}[h]{0.49\linewidth}
		\center{\includegraphics[width=\textwidth]{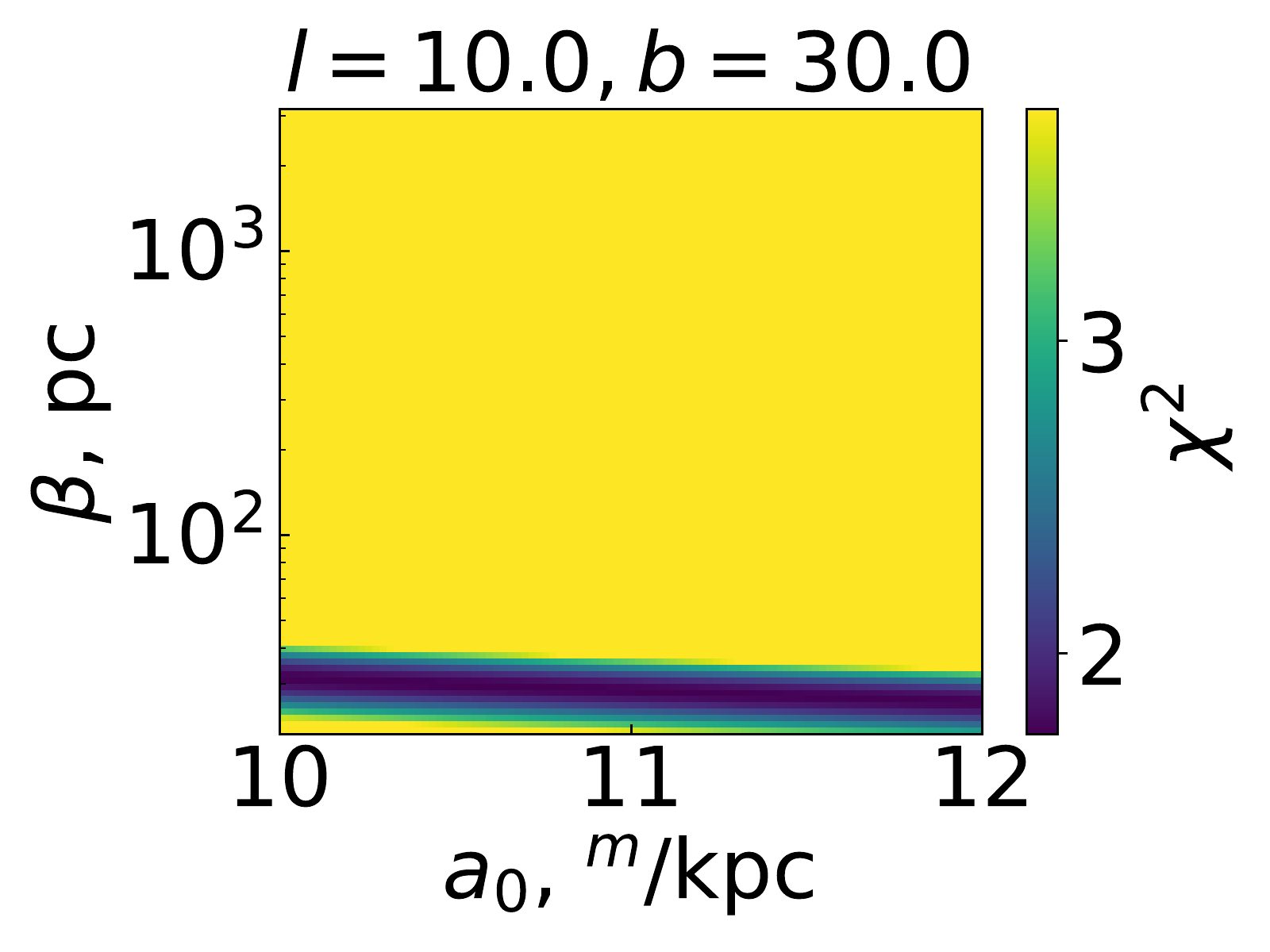}}
	\end{minipage}
	\hfill
	\begin{minipage}[h]{0.49\linewidth}
		\center{\includegraphics[width=\textwidth]{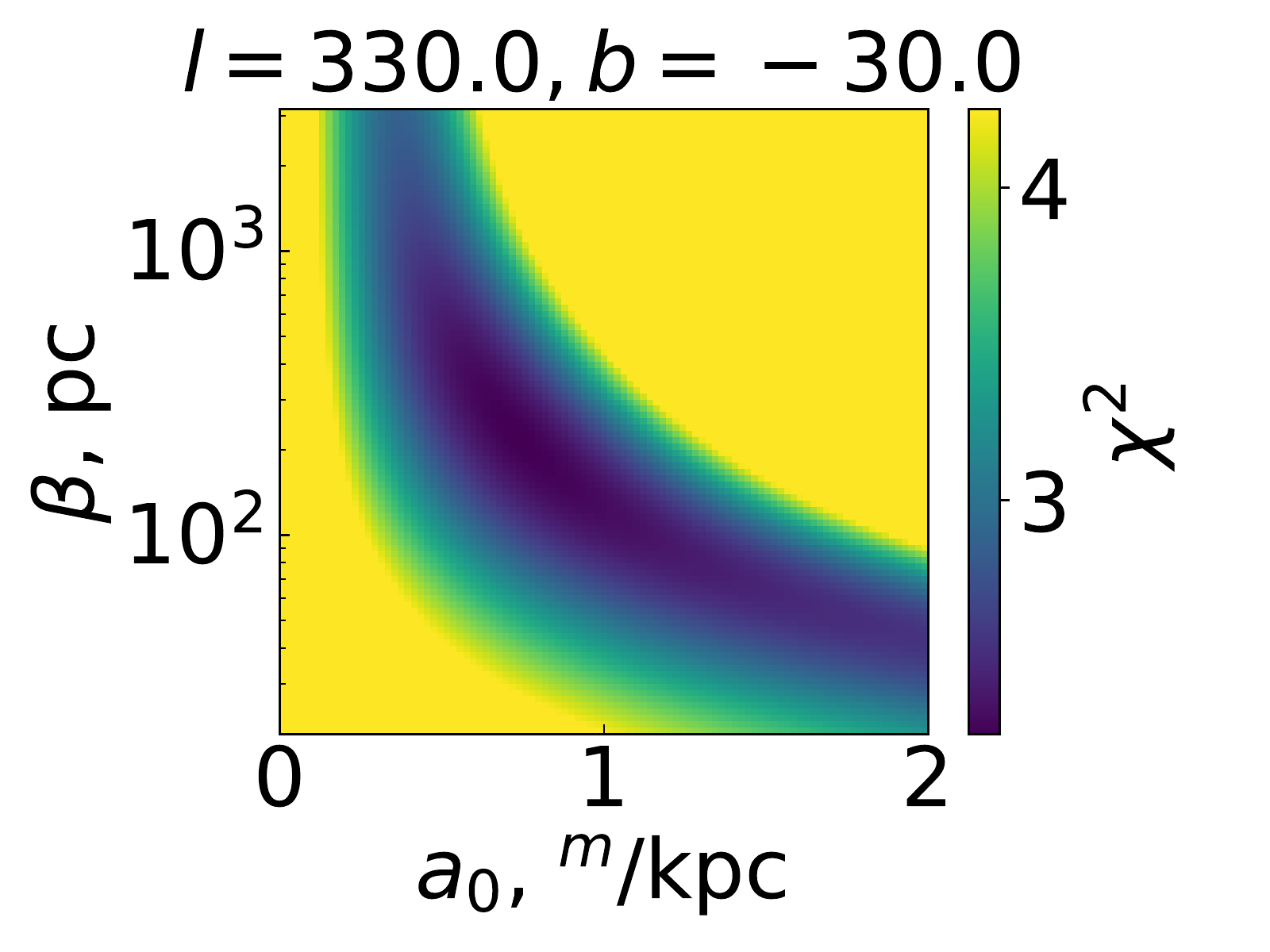}}
	\end{minipage}
	\vfill
	\begin{minipage}[h]{0.49\linewidth}
		\center{\includegraphics[width=\textwidth]{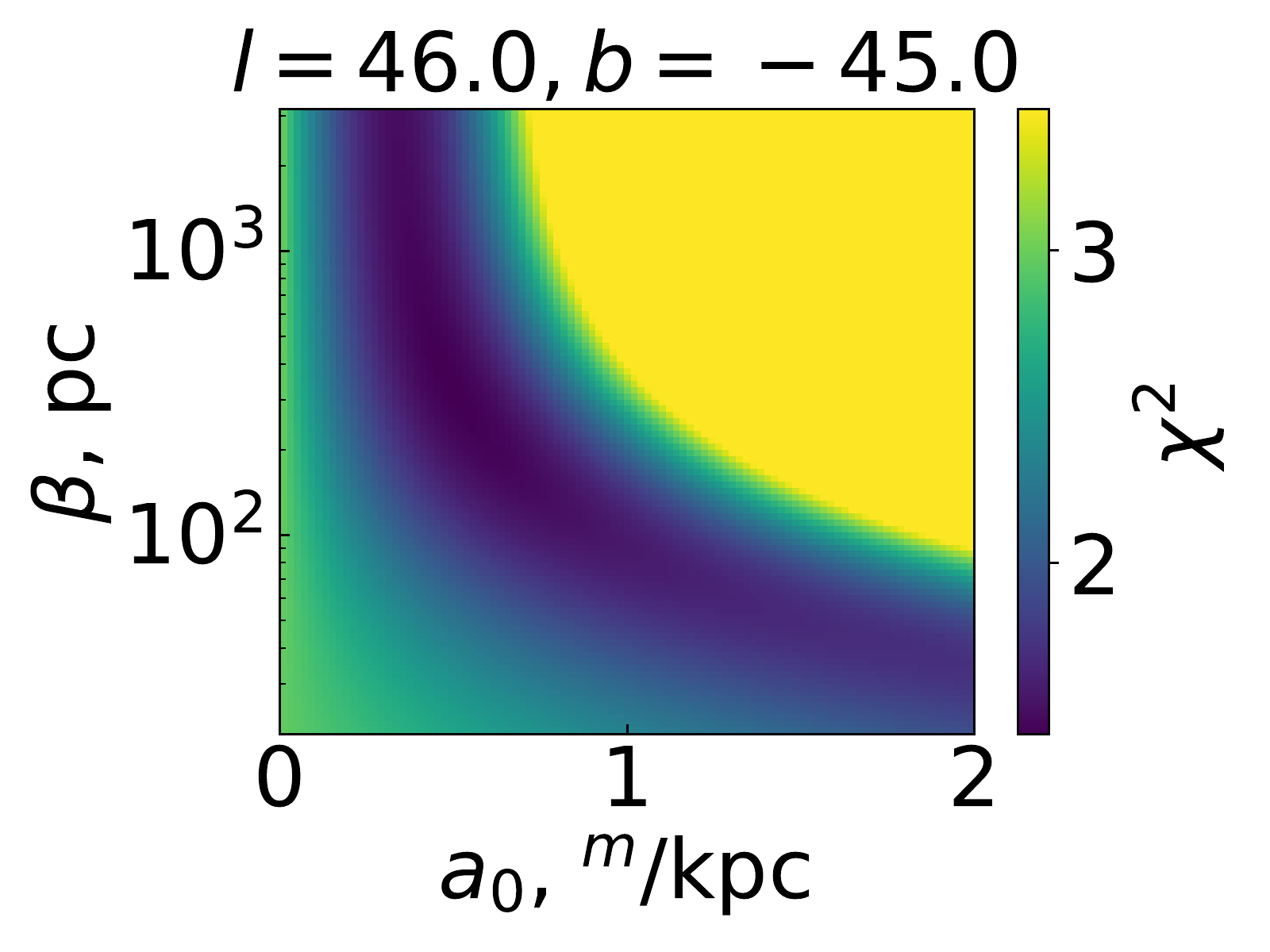}}
	\end{minipage}
	\hfill
	\begin{minipage}[h]{0.49\linewidth}
		\center{\includegraphics[width=\textwidth]{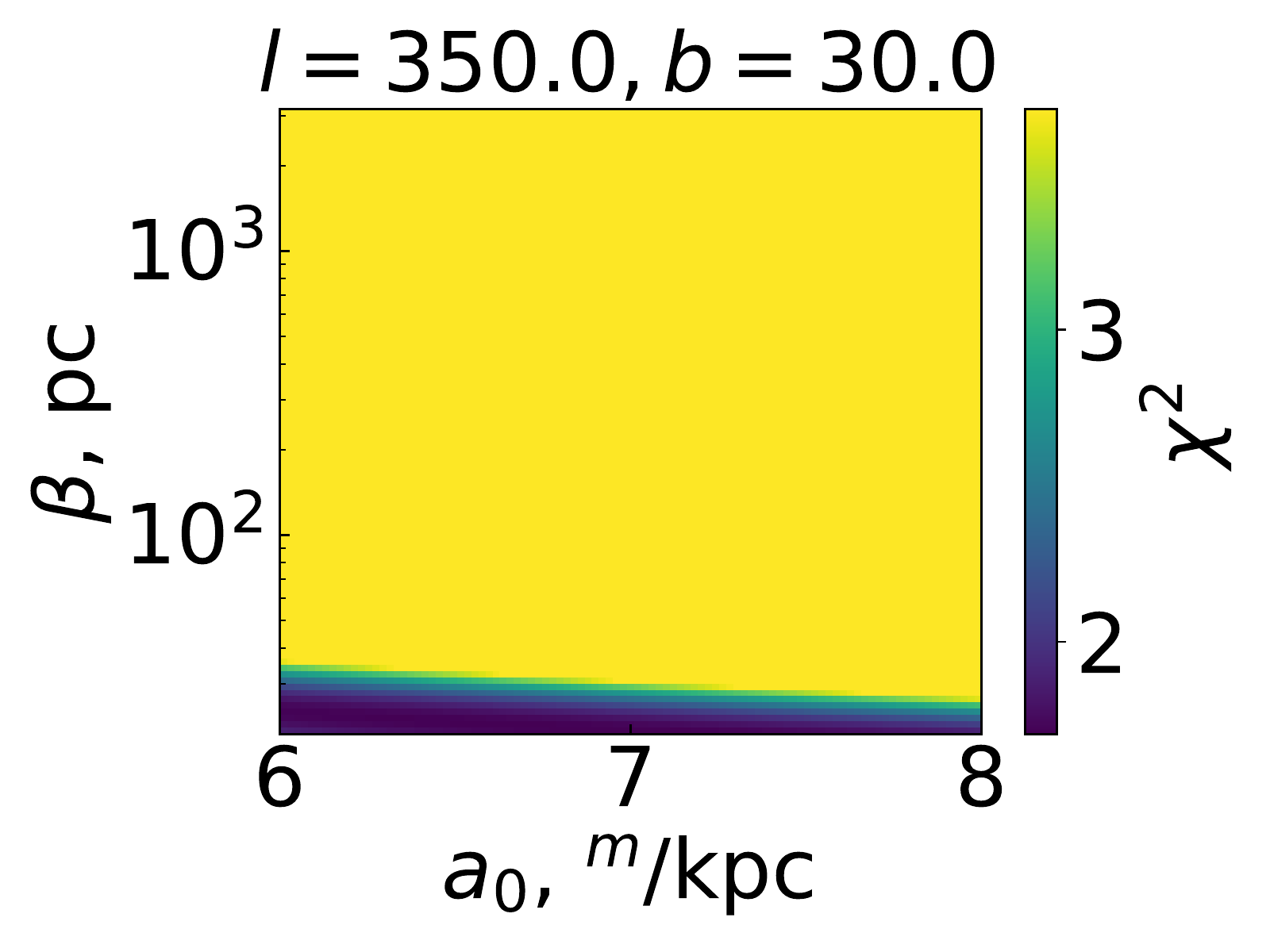}}
	\end{minipage}
	\caption{Examples of $\chi^2$-scan solutions for the same areas as in Fig.\ref{fig:apprx1}.}
    \label{fig:apprx2}
\end{figure}

Additionally to Levenberg-Marquardt algorithm of minimization we calculated the $\chi^2$ values on the ($a_0$, $\beta$) grid and found the $\chi^2$ minimum via brute-force search. We have calculated the standard errors as a $1\sigma$- contour. Fig.\ref{fig:apprx2} shows a $\chi^2$ maps for the same areas on the sky as in Fig.\ref{fig:apprx1}. The results are also indicated in Fig.\ref{fig:apprx1} as red lines. If there are only one green line visible on the plot, then two solutions coincide. \\
Size and shape of blue area on the $\chi^2$ map represents the degeneracy of the parameters, i.e the range of the parameters $a_0$ and $\beta$, corresponding to the $\chi^2$ values close to the minimum, could be quite broad. That results in large standard errors. Note that in this method errors are naturally restricted by the size of the grid and not always represent the actual uncertainty. \\
Large blue regions on the $\chi^2$ maps can also be associated with the data reliability, discussed in previous subsection.

\subsection{The results of approximation within the areas}

In both methods, the $\chi^2$ is calculated. We chose between the two methods in each area the method which provide the minimal $\chi^2$ value. Final results are presented in Tab. \ref{tab:areas}.  \\
We have also calculated the estimation for the total Galactic extinction. The total Galactic visual extinction $A_{Gal}$, i.e the extinction in V band, produced by the Galactic material, is of particular interest in the context of extragalactic researches. Moreover, it plays crucial role in distance scaling, because the scale relies on the distance to reference objects and depends on their absolute magnitudes, which require knowledge of the total Galactic extinction. \\
The Galactic extinction could be estimated from the Parenago formula (\ref{eq:parenago}) if distance tends to infinity:

\begin{equation}
    A_{Gal}(l,b) = \frac{a_0 \cdot \beta}{\sin|b|} 
\end{equation}
We have calculated values of $A_{Gal}$ for each of our areas. The results are also presented in Tab.\ref{tab:areas}. 

\begin{table*}
\begin{center}
\caption{Interstellar extinction in the selected areas}
\begin{tabular}{r|rr|lll}
Area     & l      & b          & $a_0$, $mag/kpc$               & $\beta$, $pc$                  & $A_{Gal}$ \\
\noalign{\smallskip}\hline\noalign{\smallskip}
  1   & 5.0  & -30.0      & 6  $\pm$ 16   & 14   $\pm$ 32   & 0.18  $\pm$ 0.66  \\
  2   & 30.0  & -30.0      & 0.8 $\pm$ 0.4   & 55    $\pm$ 32   & 0.09  $\pm$ 0.03  \\
  3   & 230.0  & -30.0      & 7 $\pm$ 9e5   & 3    $\pm$ 4e5   & 0.05  $\pm$ 900  \\  
  5  & 280.0  & -30.0       & 5 $\pm$ 13   & 15   $\pm$ 37   & 0.16  $\pm$ 0.56  \\
  6  & 305.0  & -30.0       & 15 $\pm$ 54   & 11   $\pm$ 41   & 0.35  $\pm$ 1.75  \\
  7  & 330.0  & -30.0       & 0.7 $\pm$ 0.2   & 216   $\pm$ 110   & 0.31  $\pm$ 0.18  \\
  8  & 17.0  & -45.0      & 3.4 $\pm$ 31.4   & 19   $\pm$ 172   & 0.09 $\pm$ 1.17  \\  
  9  & 46.0  & -45.0       & 0.5 $\pm$ 0.2   & 330   $\pm$ 397   & 0.23 $\pm$ 0.29  \\
  10  & 195.0  & -45.0       & 3.5 $\pm$ 8.5   & 32   $\pm$ 79   & 0.16 $\pm$ 0.55  \\
  12  & 270.0  & -50.0       & 0.22 $\pm$ 0.24   & 210   $\pm$ 440   & 0.06 $\pm$ 0.14  \\  
  13  & 290.0  & -45.0       & 0.9 $\pm$ 2.4   & 40   $\pm$ 115   & 0.05 $\pm$ 0.19  \\    
  15  & 335.0  & -45.0       & 0.2 $\pm$ 0.9   & 70   $\pm$ 360   & 0.02 $\pm$ 0.14  \\ 
  16  & 26.0  & -56.9       & 0.05 $\pm$ 2.2   & 50   $\pm$ 2000   & 0.003 $\pm$ 0.18  \\   
  17  & 96.7  & -72.9       & 1.6 $\pm$ 2.9   & 60   $\pm$ 120   & 0.1 $\pm$ 0.3  \\   
  18  & 200.0  & -60.0       & 2 $\pm$ 15   & 30   $\pm$ 220   & 0.08 $\pm$ 0.7  \\   
  19  & 290.0  & -60.0       & 5 $\pm$ 520   & 12   $\pm$ 1300   & 0.1 $\pm$ 10  \\ 
  20  & 20.0  & -70.0       & 6 $\pm$ 7e4   & 7   $\pm$ 8e4   & 0.05 $\pm$ 800  \\   
  22  & 340.0  & -70.0       & 0.21 $\pm$ 0.04   & 190   $\pm$ 700   & 0.04 $\pm$ 0.17  \\ 
  23  & 275.0  & 30.0       & 3.2 $\pm$ 8.5   & 15   $\pm$ 43   & 0.1 $\pm$ 0.4  \\ 
  24  & 290.0  & 30.0       & 0.9 $\pm$ 0.5   & 115   $\pm$ 114   & 0.2 $\pm$ 0.2  \\ 
  25  & 320.0  & 30.0       & 6 $\pm$ 390   & 9   $\pm$ 590   & 0.1 $\pm$ 9.5  \\ 
  26  & 350.0  & 30.0       & 9.2 $\pm$ 8.1   & 15   $\pm$ 13   & 0.28 $\pm$ 0.35  \\  
  27  & 10.0  & 30.0       & 10.5 $\pm$ 2.2   & 29.3   $\pm$ 6.6   & 0.61 $\pm$ 0.19  \\   
  28  & 113.0  & -85.0       & 0.1 $\pm$ 0.5   & 140  $\pm$ 1000   & 0.014 $\pm$ 0.119  \\  
  29  & 260.3  & 46.3       & 0.5 $\pm$ 0.9   & 100  $\pm$ 200   & 0.07 $\pm$ 0.19  \\  
  30  & 279.5  & 45.8       & 0.42 $\pm$ 0.45   & 110  $\pm$ 170   & 0.07 $\pm$ 0.12  \\ 
  31  & 305.0  & 45.0       & 6 $\pm$ 30   & 20  $\pm$ 100   & 0.17 $\pm$ 1.25 \\ 
  32  & 330.0  & 45.0       & 2.50 $\pm$ 1.15   & 55  $\pm$ 28   & 0.19 $\pm$ 0.13 \\ 
  33  & 242.0  & -22.0       & 2.3 $\pm$ 25.0   & 10  $\pm$ 110   & 0.06 $\pm$ 0.92 \\
  34  & 269.0  & -23.0       & 6.4 $\pm$ 21.0   & 13  $\pm$ 45   & 0.2 $\pm$ 1.0 \\
  35  & 310.0  & -22.0       & 4.8 $\pm$ 1.6   & 41  $\pm$ 19   & 0.5 $\pm$ 0.3 \\
  36  & 19.9  & -24.8       & 6.9 $\pm$ 3.5   & 28  $\pm$ 16   & 0.5 $\pm$ 0.4 \\
  37  & 262.3  & 22.3       & 19 $\pm$ 7e6   & 1.5  $\pm$ 5e5   & 0.07 $\pm$ 4e5 \\  
  38  & 335.0  & -26.5       & 7.5 $\pm$ 2.6   & 22  $\pm$ 8   & 0.37 $\pm$ 0.19 \\  
  39  & 283.9  & -22.4       & 1.91 $\pm$ 1.98   & 31  $\pm$ 39   & 0.16 $\pm$ 0.25 \\    
  40  & 245.4  & -20.3       & 4.3 $\pm$ 7.6   & 20  $\pm$ 38   & 0.26 $\pm$ 0.68 \\

\end{tabular}
\label{tab:areas}
\end{center}
\end{table*}

\section{Approximation of cosecant law parameters over the entire sky and final formula}

\begin{figure*}
    \centering
    \includegraphics[width=0.9\linewidth]{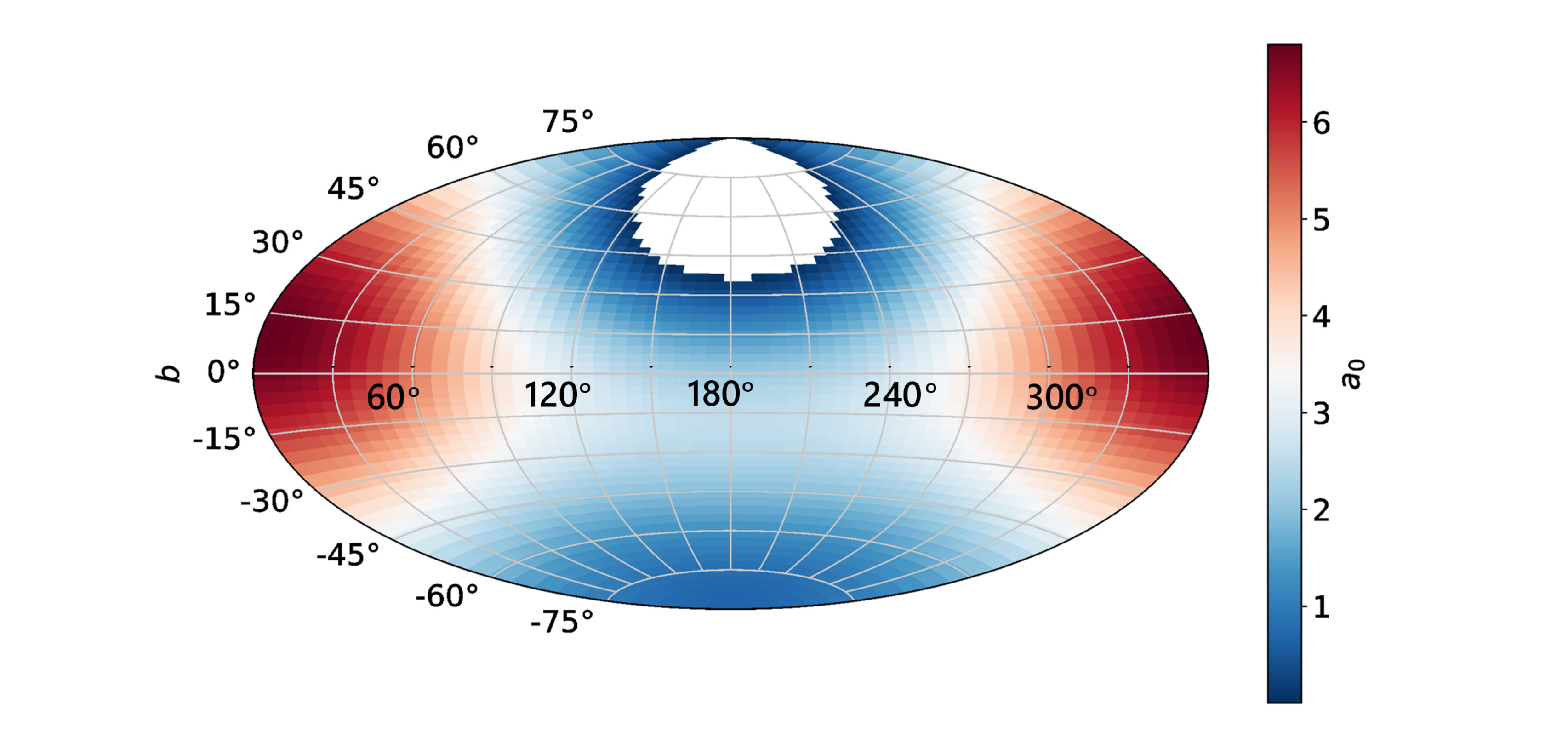} \\
     \includegraphics[width=0.93\linewidth]{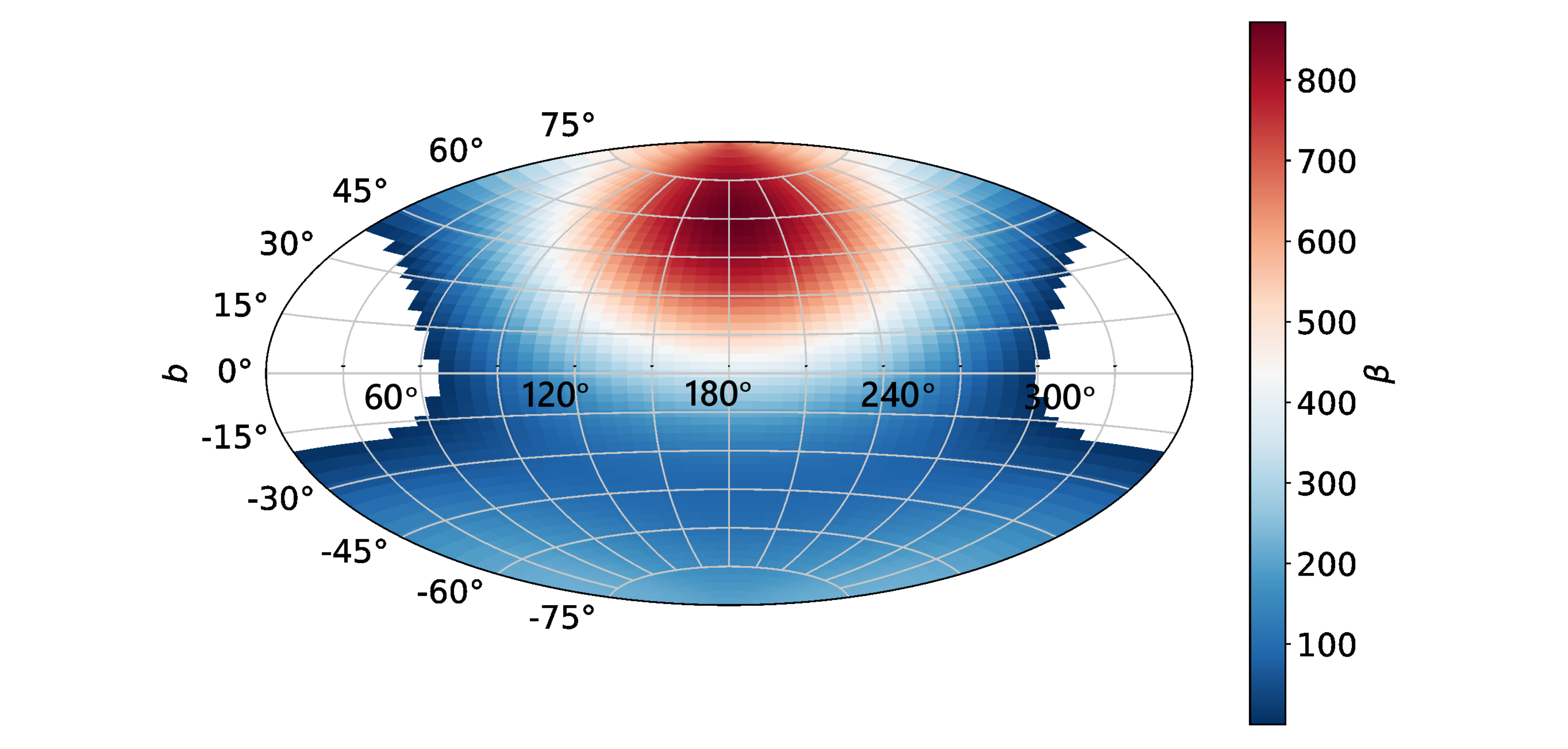} \\
    \includegraphics[width=0.9\linewidth]{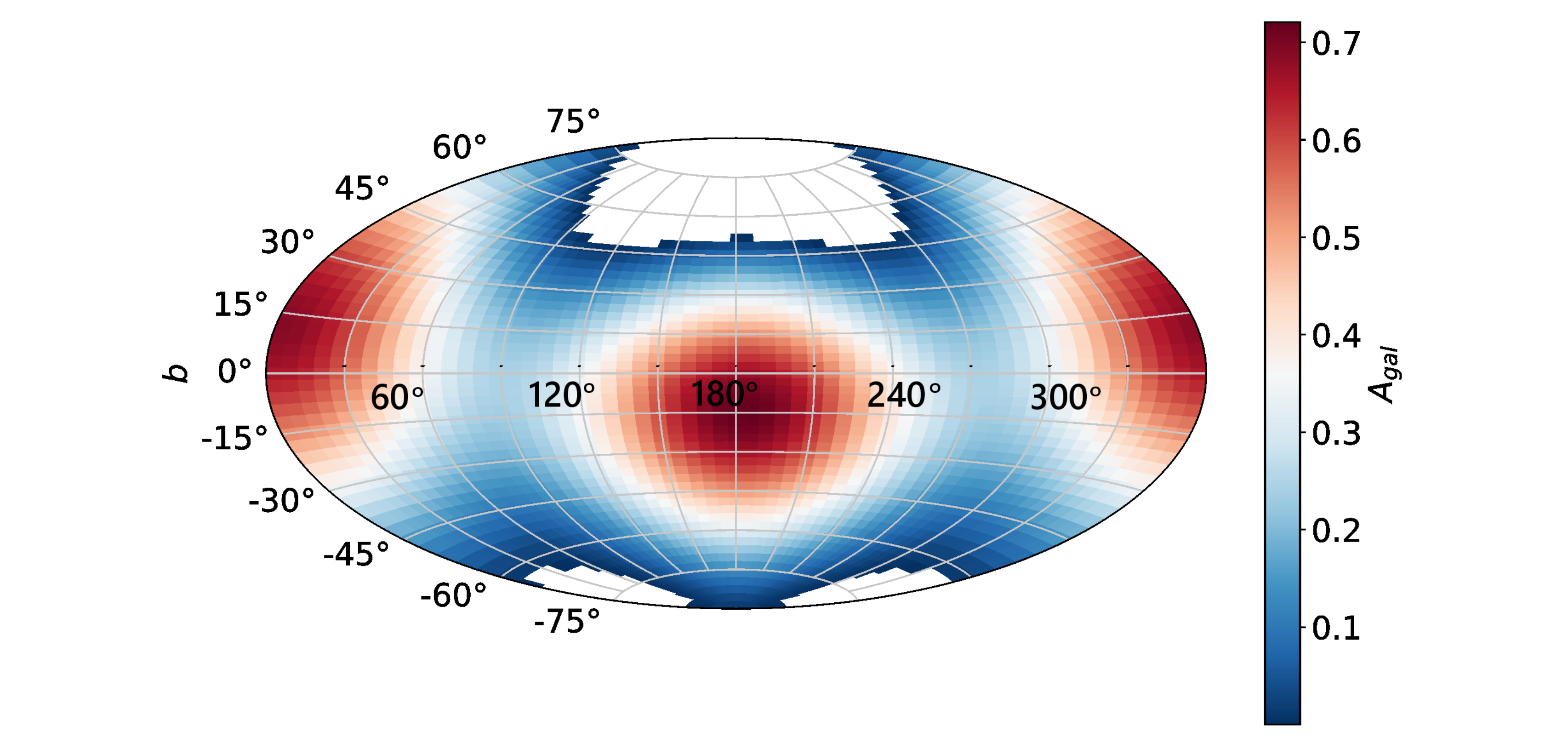} \\
    \caption{Approximation of parameters over the sky. The areas with negative values are indicated as blank spaces.} 
    \label{fig:fig5}
\end{figure*}

Finally, we have approximated the parameters $a_0$, $\beta$ and the total Galactic extinction $A_{Gal}$ by a polynomial composed of spherical harmonics of degree and order 2:
\begin{eqnarray} \label{eq:harm}
    f(l,b) = A_{00} Y_0^0 \left(l, \frac{\pi}{2} - b\right) + A_{10} Y_1^0 \left(l, \frac{\pi}{2} - b\right) +\nonumber \\+ A_{11} Y_1^1 \left(l, \frac{\pi}{2} - b\right) + A_{20} Y_2^0 \left(l, \frac{\pi}{2} - b\right) +\nonumber \\ + A_{21} Y_2^1 \left(l, \frac{\pi}{2} - b\right) + A_{22} Y_2^2 \left(l, \frac{\pi}{2} - b\right)
\end{eqnarray}
For the approximation, we used the parameters for 36 out of 40 areas. They are listed in Tab \ref{tab:areas}. We also used the data obtained from  \cite{2021EPJST.tmp..185N}. The coefficient of the polynomial for $a_0$, $\beta$ and $A_{Gal}$ approximation are presented in Tab.\ref{tab:spher}. 
\begin{table}
\begin{center}
    \caption{The coefficients of the polynomial approximation for $a_0$, $\beta$ and $A_{Gal}$.}
\begin{tabular}{l|lll}
       $f(l,b)$    & $a_0$             & $\beta$ & $A_{Gal}$  \\
    \noalign{\smallskip}\hline\noalign{\smallskip}
    $A_{00}$ &  10.0 $\pm$ 1.5  &  811.4 $\pm$ 210.2  &  0.967 $\pm$ 0.113     \\
    $A_{10}$ & -0.9 $\pm$ 1.5 &  495.3 $\pm$  212.4 & -0.148 $\pm$ 0.114 \\
    $A_{11}$ & -6.64 $\pm$ 2.18  & 715.7  $\pm$ 298.9  & 0.02 $\pm$ 0.16 \\
    $A_{20}$ & -4.0 $\pm$ 1.8 &  336.3 $\pm$  250.6 &  -0.51 $\pm$ 0.13 \\
    $A_{21}$ & -2.6 $\pm$ 1.9 &  520.2 $\pm$ 266.8  & -0.43  $\pm$ 0.14  \\
    $A_{22}$ & 0.77 $\pm$ 2.19 &  32.4 $\pm$ 300.5  &  0.56 $\pm$ 0.16    \\

    \end{tabular}
    \label{tab:spher}
\end{center}
\end{table}
Substituting the values of the coefficients $A_{ij}$ into the Eq.(\ref{eq:harm}) and then into the cosecant law (Eq.\ref{eq:parenago}), one could get an analytical expression to estimate the interstellar extinction in particular line of sight at particular distance. \\
We should note, that due to the harmonic form of the function, there are areas in approximation where the values of parameters become negative. They are shown in Fig. \ref{fig:fig5} with blank spaces. In these areas we either can say nothing about the extinction or it is approximately equals to zero (as on polar region). We do not recommend to use the approximation within these areas. We also want to emphasise, that we have not used areas lower that $|b|<20^\circ$, so we kindly ask to be careful using the resulting formula lower than this border.  

\subsection{Uncertainties}
\begin{figure}
    \centering
    \includegraphics[width=\columnwidth]{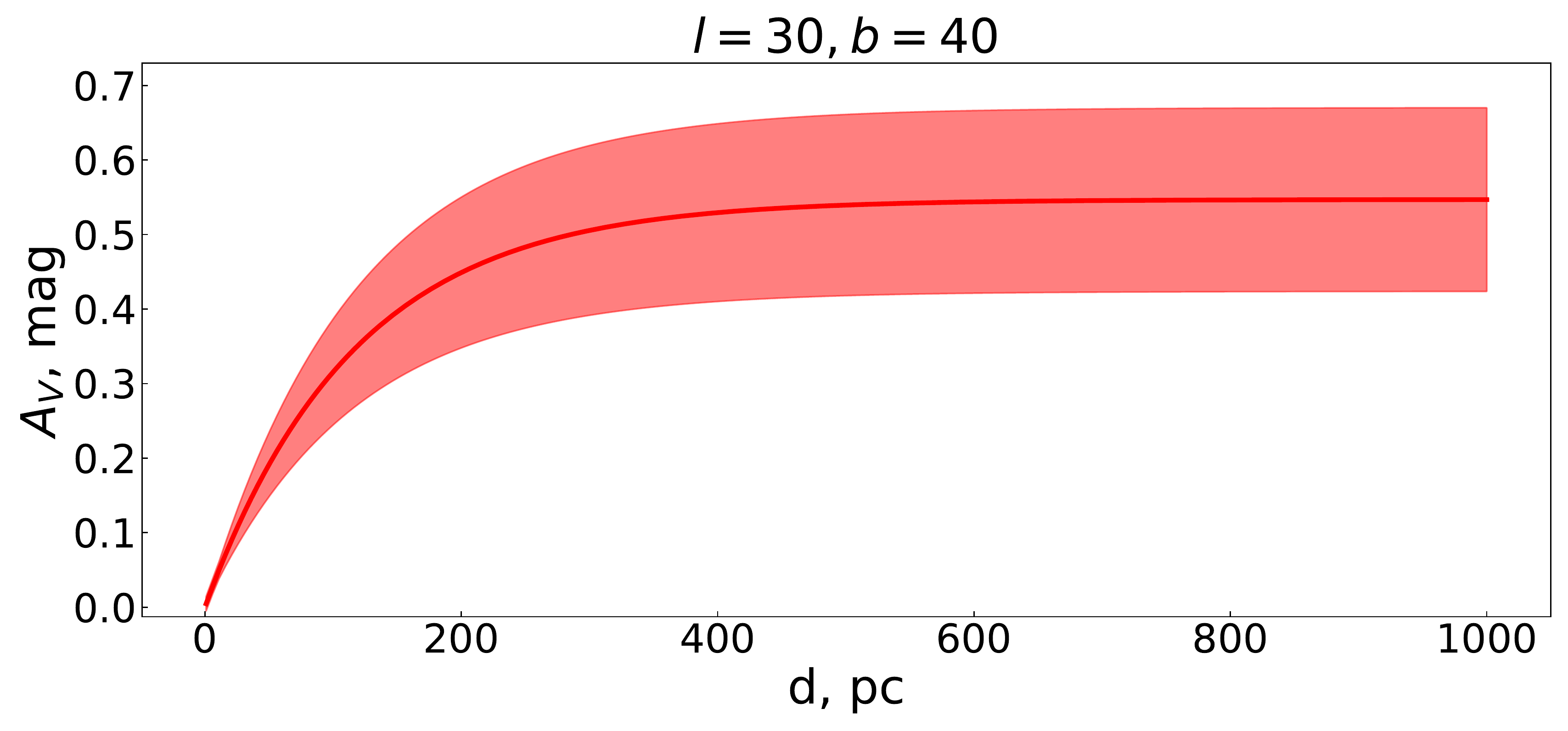} \\
    \includegraphics[width=\columnwidth]{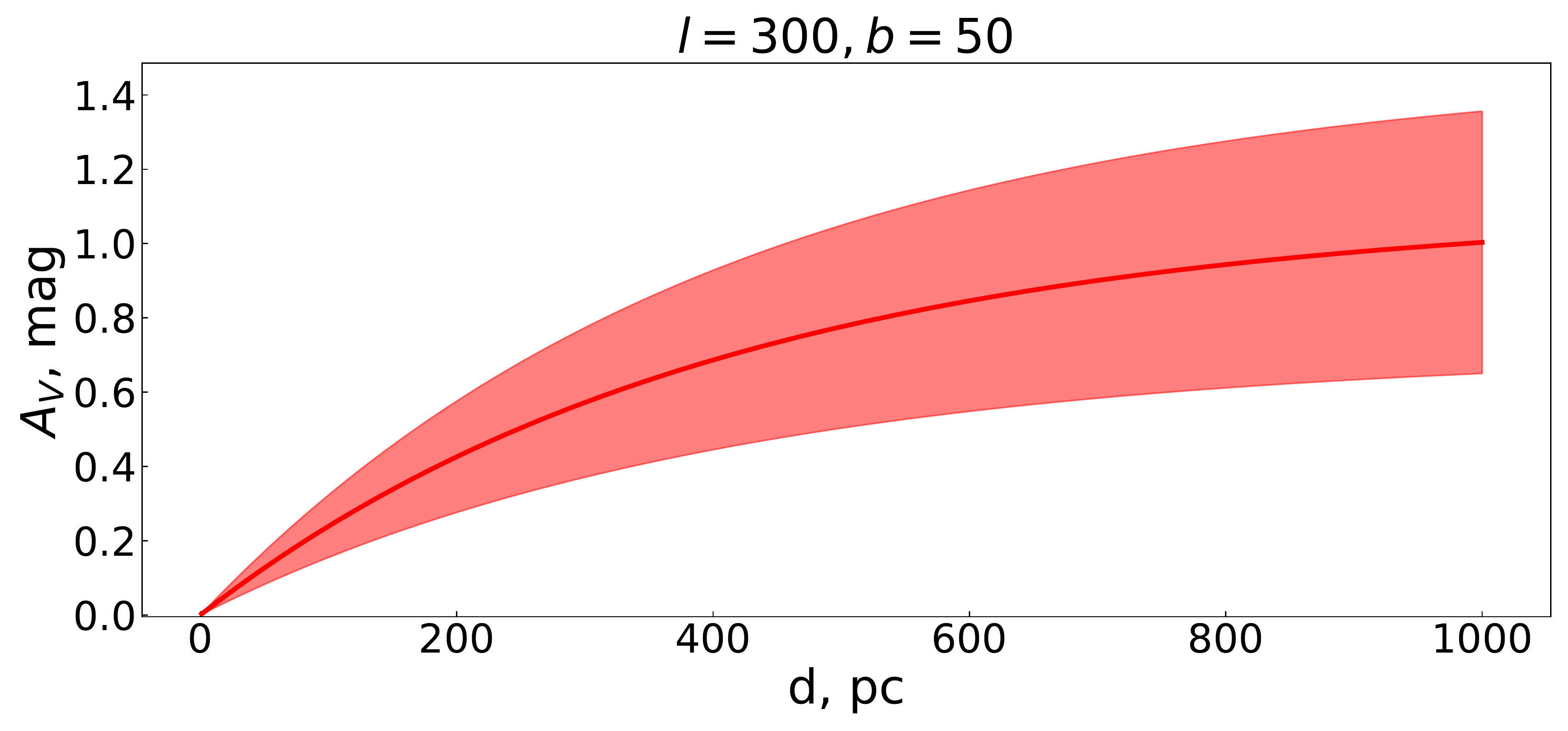} \\
    \includegraphics[width=\columnwidth]{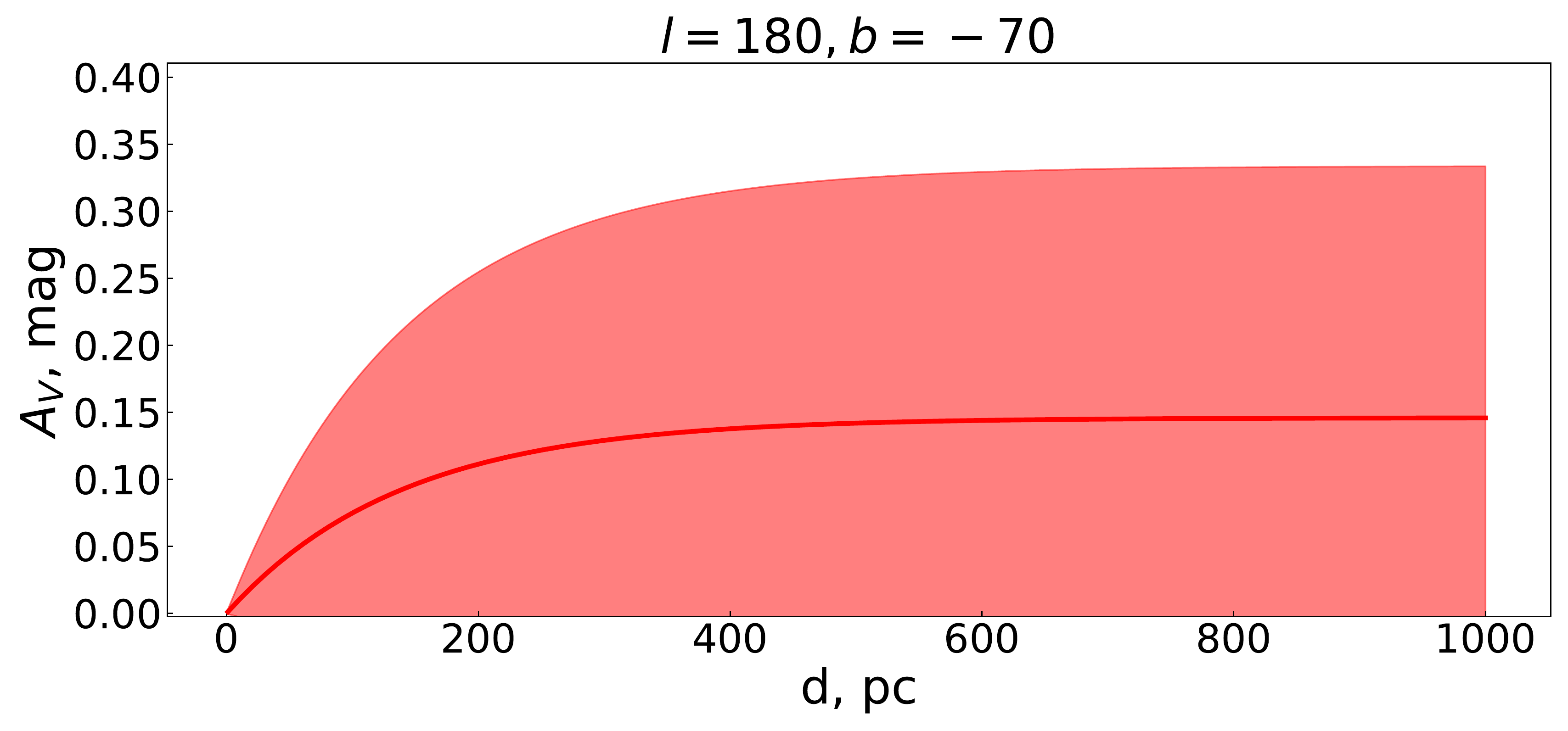} \\
    \caption{The examples of the visual extinction to distance dependence, calculated with our formula (Eq.\ref{eq:harm}). The uncertainties were also calculated and shown as a colour-filled area.} 
    \label{fig:fig6}
\end{figure}

The uncertainties of  $a_0$, $\beta$ and $A_{Gal}$ depend on latitude and longitude of the line of sight and can be calculated as follows:

\begin{eqnarray}
    \delta f(l,b) = 
   \bigg[ \delta A_{00}^2\cdot k_{00}^2 + \delta A_{10}^2\cdot k_{10}^2 + \nonumber \\
    \delta A_{11}^2\cdot k_{11}^2 + \delta A_{20}^2\cdot k_{20}^2 + \nonumber \\ 
    \delta A_{21}^2\cdot k_{21}^2 + \delta A_{22}^2\cdot k_{22}^2 \bigg] ^{1/2}
\end{eqnarray}
where $k_{i,j} = Y_i^j (l, \frac{\pi}{2} - b)$. \\
One can also estimate the relative uncertainty of $A_V$ with this formula which depends on latitude, longitude and distance:

\begin{eqnarray}
    \frac{\delta A_V}{A_V} (l,b,d) =  \bigg[ \frac{\delta a_0^2}{a_0^2} + \frac{\delta \beta^2}{\beta^2} \cdot \bigg(1  + \frac{d\cdot \sin(|b|)}{\beta} \nonumber \cdot \\ \cdot \frac{\exp(\frac{-d\cdot \sin(|b|)}{\beta})}{(1-\exp(\frac{-d\cdot \sin(|b|)}{\beta})} \bigg) \bigg]^{1/2} 
\end{eqnarray}
The examples of the interstellar extinction $A_v$ calculated with the uncertainty are shown in Fig.\ref{fig:fig6}. Generally, the relative uncertainty tends to be larger on higher latitudes, although it depends on the longitude as well.   

\section{Summary and future plans}
We have obtained the interstellar visual extinction of objects in 40 areas of the southern sky with Gaia EDR3 and RAVE DR6 data. Comparison of calculated extinctions with various interstellar extinction models or maps shows qualitative similarity, which indicates the adequacy of chosen method of extinction determination. \\     
Approximation in areas showed that the quality of data is not satisfying. In some areas no successful approximation has been performed due to odd trend. We assume that this problem is caused by inadequate RAVE temperatures. Hopefully, homogenisation will be made and eliminate that problem. \\
The all-sky approximation with spherical harmonics was made and the approximate analytical description was obtained. Nonetheless, there are regions where the solution tends to zero within the margin of error (regions with negative values of parameters), so no reasonable solution can be found. It is necessary to reduce the approximation error by an order of magnitude to obtain reasonable solutions for areas with low total extinction. Additionally, we do not recommend to apply the formula to the latitudes below 20 degrees north or south, since we have not utilised the areas close to the Galactic plane (we assume that the cosecant law is not suitable there). \\ 
The issue with the discrepancy of effective temperatures adopted from LAMOST and RAVE surveys still remains. Using other surveys (like APOGEE or SEGUE) will probably help us to get an additional information and draw the conclusions about the reliability of the data used by comparing temperatures.\\
We also plan to include stars from the outside of main sequence both in Southern and Northern sky and extend our investigation to low galactic latitudes probably with more complex $A_V(d)$ law. 

\begin{acknowledgement}

Authors thank Kirill Grishin for fruitful discussion and help with approximation methods. This work was partly supported by the RFBR grant 20-52-53009. This research has made use of NASA's Astrophysics Data System and VizieR catalogue access tool, CDS, Strasbourg, France. This work has made use of data from the European Space Agency (ESA) mission
{\it Gaia} (\url{https://www.cosmos.esa.int/gaia}), processed by the {\it Gaia}
Data Processing and Analysis Consortium (DPAC,
\url{https://www.cosmos.esa.int/web/gaia/dpac/consortium}). Funding for the DPAC
has been provided by national institutions, in particular the institutions
participating in the {\it Gaia} Multilateral Agreement.

\end{acknowledgement}

\bibliographystyle{apalike}
\bibliography{ext}

\end{document}